\documentclass[twocolumn]{aastex631}
\usepackage{gensymb}

\usepackage{soul}
\newcommand{\loicadd}[1]{\textcolor{black}{#1}}

\begin{document}

\title{JWST 1.5~$\mu$m and 4.8~$\mu$m Photometry of Y Dwarfs}

\correspondingauthor{Lo\"ic Albert}
\email{loic.albert@umontreal.ca}

\author[0000-0003-0475-9375]{Lo\"ic Albert$^{\star}$}
\affiliation{D\'epartement de Physique, Observatoire du Mont-M\'egantic and Trottier Institute for Research on Exoplanets, Universit\'e de Montr\'eal, C.P. 6128, Succ. Centre-ville, Montr\'eal, H3C 3J7, Québec, Canada}
\affiliation{Institut Trottier de Recherche sur les exoplan\`etes, Universit\'e de Montr\'eal}
\author[0000-0002-3681-2989]{Sandy K. Leggett$^{\star}$}
\affil{NOIRLab - Gemini North (HI)}
\author[0000-0002-5335-0616]{Per Calissendorff}
\affil{Department of Astronomy, University of Michigan, Ann Arbor, MI 48109, USA}
\author[0000-0002-5922-8267]{Thomas Vandal}
\affiliation{D\'epartement de Physique and Observatoire du Mont-M\'egantic, Universit\'e de Montr\'eal, C.P. 6128, Succ. Centre-ville, Montr\'eal, H3C 3J7, Québec, Canada.}
\affiliation{Institut Trottier de Recherche sur les exoplan\`etes, Universit\'e de Montr\'eal}
\author[0000-0003-4269-260X]{J. Davy Kirkpatrick}
\affil{California Institute of Technology}
\author[0000-0001-8170-7072]{Daniella C. Bardalez Gagliuffi}
\affil{Department of Physics \& Astronomy, Amherst College, 25 East Drive, Amherst, MA 01003, USA}
\affil{Department of Astrophysics, American Museum of Natural History, 200 Central Park West, New York, NY 10024, USA}
\author[0000-0003-1863-4960]{Matthew De Furio}
\affil{Department of Astronomy, University of Texas at Austin, Austin, TX 78712, USA}

\author[0000-0003-1227-3084]{Michael Meyer}
\affil{Department of Astronomy, University of Michigan, Ann Arbor, MI 48109, USA}

\author[0000-0002-5627-5471]{Charles A. Beichman}
\affil{Jet Propulsion Laboratory}
\author[0000-0002-6523-9536]{Adam J. Burgasser}
\affil{University of California, San Diego, La Jolla, CA, USA}
\author[0000-0001-7780-3352]{Michael C. Cushing}
\affil{Ritter Astrophysical Research Center, Department of Physics and Astronomy, University of Toledo, 2801 W. Bancroft Street, Toledo, OH 43606, USA}
\author[0000-0001-6251-0573]{Jacqueline Kelly Faherty}
\affil{Astrophysics Department, American Museum of Natural History, 79th Street at Central Park West, New York, NY 10024}
\author[0000-0002-2428-9932]{Cl\'emence Fontanive}
\affiliation{D\'epartement de Physique and Observatoire du Mont-M\'egantic, Universit\'e de Montr\'eal, C.P. 6128, Succ. Centre-ville, Montr\'eal, H3C 3J7, Québec, Canada.}
\affiliation{Institut Trottier de Recherche sur les exoplan\`etes, Universit\'e de Montr\'eal}
\author[0000-0001-5072-4574]{Christopher R. Gelino}
\affil{California Institute of Technology}
\author[0000-0002-8916-1972]{John E. Gizis}
\affil{University of Delaware}
\author[0000-0002-7162-8036]{Alexandra Z. Greenbaum}
\affiliation{IPAC, Mail Code 100-22, Caltech, 1200 E. California Blvd., Pasadena, CA 91125, USA}

\author[0000-0003-1180-4138]{Frantz Martinache}
\affil{Université Côte d’Azur, Observatoire de la Côte d'Azur, CNRS, Laboratoire Lagrange, France}

\author[0000-0002-1721-3294]{Mamadou N'Diaye}
\affil{Université Côte d’Azur, Observatoire de la Côte d'Azur, CNRS, Laboratoire Lagrange, France}
\author[0000-0003-2595-9114]{Benjamin J. S. Pope}
\affiliation{School of Mathematics and Physics, The University of Queensland, St Lucia, QLD 4072, Australia}
\affiliation{Centre for Astrophysics, University of Southern Queensland, West Street, Toowoomba, QLD 4350, Australia}
\author[0000-0002-6730-5410]{Thomas L. Roellig}
\affil{MS 245-6, NASA Ames Research Center, Moffett Field, CA 94035}
\author[0000-0001-9525-3673]{Johannes Sahlmann}
\affil{European Space Agency (ESA), European Space Astronomy Centre (ESAC), Camino Bajo del Castillo s/n, 28692 Villanueva de la Ca\~nada, Madrid, Spain}

\author[0000-0003-1251-4124]{Anand Sivaramakrishnan}
\affiliation{Astrophysics Department, American Museum of Natural History, 79th Street at Central Park West, New York, NY 10024}
\affiliation{Space Telescope Science Institute, 3700 San Martin Drive, Baltimore, MD 21218, USA}
\affiliation{Department of Physics and Astronomy, Johns Hopkins University, 3701 San Martin Drive, Baltimore, MD 21218, USA}

\author[0000-0001-7591-2731]{Marie Ygouf}
\affil{Jet Propulsion Laboratory}

\begin{abstract}

Brown dwarfs lack nuclear fusion and cool with time; the coldest known have an effective temperature below 500\,K, and are known as Y dwarfs. We present  a {\em James Webb Space Telescope (JWST)} photometric dataset of Y dwarfs: twenty-three were imaged in wide-field mode, 20 using NIRCam with the F150W and F480M filters, and 3 using NIRISS with the F480M filter. We present an F480M vs. F150W $-$ F480M color-magnitude diagram for our sample, and other brown dwarfs with F150W and F480M colors synthesized from {\it JWST} spectra by \citet{Beiler_2024}. For one target, WISEA\,J083011.95$+$283716.0, its detection in the near-infrared confirms it as one of the reddest Y dwarfs known, with F150W $-$ F480M $= 9.62$\, mag. We provide its updated parallax and proper motion. One of the Beiler et al. Y dwarfs, CWISEP J104756.81+545741.6, is unusually blue, consistent with strong CO absorption seen in its spectrum which the F480M filter is particularly sensitive to.  The strong CO and the kinematics of the object suggest it may be very low-mass and young. We update the resolved photometry for the close  binary system WISE\,J033605.05$-$014350.4 AB, and find that the secondary is almost as cold as WISE\,085510.83$-$071442.5, with $T_{\rm eff} \lesssim 300$~K, however the F150W $-$ F480M color is significantly bluer, possibly suggesting the presence of water clouds. Astrometry is measured at the {\it JWST} epoch for the sample which is consistent with parallax and proper motion values reported by \cite{Kirkpatrick_2021} and Marocco et al. (in prep).  

\end{abstract}


\section{Introduction} \label{sec:intro}

Y dwarfs are the coolest known class of brown dwarfs - objects with insufficient mass for sustained nuclear fusion at their core \citep{Burrows_2001}. The Y dwarfs have effective temperatures ($T_{\rm eff}$) less than 500~K \citep{Kirkpatrick_2021}, and the coldest Y dwarf currently known, WISE\,085510.83$-$071442.5, has $T_{\rm eff} \approx 260$~K \citep[see also Section 5]{Leggett_2021, Luhman_2024}. Evolutionary models \citep{Marley_2021} calculate that field Y dwarfs, with ages of a few Gyr \citep[e.g.][]{Kirkpatrick_2021, Best_2024}, have masses of about 10 Jupiter-masses ($M_{\rm Jup}$), within the planetary-mass regime, such that properties of giant planets and cold brown dwarfs overlap significantly \citep{Morley_2014, Showman_2019}. Ultimately, establishing the nature of these low-mass objects, planets or brown dwarfs, needs to involve their formation pathways. To progress along that question requires a diversity of approaches: comparing and contrasting the properties of low-mass companions to host stars, e.g. orbital eccentricities \citep{Bowler_2020}, exploring the lower mass limits of star-formation \citep{Kirkpatrick_2021, kirkpatrick_2024} and characterizing the Y dwarf multiplicity to understand their formation \citep{Fontanive_2023}.

With our {\em JWST} Cycle 1 program (PID 2473, PI Lo\"ic Albert), we explore the multiplicity of 20 Y dwarfs with NIRCam kernel phase interferometry \citep{martinache.2010,ceau.2019,kammerer.2023}, using the F150W and F480M filters. In this paper we report the photometric measurements for the sample, and also provide checks on the published proper motions and parallaxes of \cite{Kirkpatrick_2021}. A companion paper (Vandal et al. in prep.) presents the results of the search for companions for the entire sample. An early discovery of the first binary Y dwarf system \loicadd{from this program} was presented by \citet{Calissendorff_2023}. Another companion paper presents detailed modeling of the photometry of WISEA~J083011.95$+$283716.0 (Matuszewska et al. in prep.). We supplement the photometry determined as part of program 2473 with measurements of cold brown dwarfs using the F480M filter, executed as part of the Cycle 1 GTO programs 1189 (PI Thomas L Roellig) and 1230 (PI Catarina Alves de Oliveira).

Section 2 describes the sample and the observations. Section 3 describes the data reduction and presents the photometric measurements. Section 4 discusses the kinematics of the sample.  The brown dwarf colors are explored using color-magnitude diagrams and models, in Section 5. Section 6 presents our conclusions. 

\bigskip
\section{Sample and Observations} \label{sec:style}

\subsection{Sample Selection}

\begin{table*}[!ht]
    \centering
            \caption{Sample of Brown Dwarfs}
            \label{tab:sample}
    \begin{tabular}{llcrcrr}
    \hline
    Short Name & Full Name & Disc. & $T_{\rm eff}$ & $T_{\rm eff}$ & Parallax & W2\\
     &  & Ref. & Est.~K & Ref. & (mas) & (Mag) \\
    \hline \hline
    \multicolumn{7}{c}{\textbf{NIRCam F150W$+$F480M Survey (Program ID 2473)}} \\
    WISE-0304  & WISE J030449.03$-$270508.3  & 1 & 465 \loicadd{± 88} & a & 73.1 ± 2.6  & 15.578 ± 0.041\\
    WISE-0336AB  & WISE J033605.05$-$014350.4AB  & 2 &  \loicadd{460 ± 79}& b & 99.8 ± 2.1  & 14.664 ± 0.023\\
    WISE-0359  & WISE J035934.06$-$540154.6  & 2 & \loicadd{$443_{-19}^{+23}$}& c & 73.6 ± 2.0  & 15.412 ± 0.026\\
    WISE-0410  & WISEPA J041022.71+150248.5  & 3 & \loicadd{451 ± 88}& a & 151.3 ± 2.0 & 14.104 ± 0.017\\
    WISE-0535  & WISE J053516.80$-$750024.9  & 2 &  \loicadd{$496_{-23}^{+28}$}& c & 68.7 ± 2.0  & 14.996 ± 0.020\\
    WISE-0647  & WISE J064723.23$-$623235.5  & 4 & \loicadd{393 ± 88}& a & 99.5 ± 1.7  & 15.115 ± 0.018\\
    WISE-0713  & WISE J071322.55$-$291751.9  & 2 &  \loicadd{464 ± 88}& a & 109.3 ± 2.1 & 14.327 ± 0.016\\
    WISE-0734  & WISE J073444.02$-$715744.0  & 2 & \loicadd{$466_{-20}^{+24}$}& c & 74.5 ± 1.7  & 15.241 ± 0.022\\
    WISE-0825  & WISE J082507.35$+$280548.5  & 5 & \loicadd{$387_{-15}^{+15}$}& c & 152.6 ± 2.0 & 14.655 ± 0.024\\
    WISE-0830  & WISEA J083011.95$+$283716.0 & 6 & \loicadd{367 ± 79}& a & 90.6 ± 13.7$^{\dagger}$ & 16.004 ± 0.072\\
    WISE-1141  & WISEA J114156.67$-$332635.5 & 7 & \loicadd{460 ± 79}& a & 104.0 ± 2.9 & 14.632 ± 0.030\\
    WISE-1206  & WISE J120604.38$+$840110.6  & 8 & \loicadd{$472_{-20}^{+26}$}& c & 84.7 ± 2.1  & 15.173 ± 0.023\\
    WISE-1405  & WISEPC J140518.40$+$553421.4& 3 & \loicadd{$392_{-15}^{+16}$}& c & 158.2 ± 2.6 & 14.098 ± 0.014\\
    WISE-1446  & CWISEP J144606.62$-$231717.8& 9 & \loicadd{$351_{-13}^{+16}$} & c &  
    \loicadd{103.8 ± 5.0}  & 15.955 ± 0.072\\
    WISE-1541  & WISEPA J154151.66$-$225025.2& 3 & \loicadd{$411_{-17}^{+18}$}& c & 166.9 ± 2.0 & 14.218 ± 0.030\\
    WISE-1639  & WISEA J163932.75$+$184049.4 & 9 & \loicadd{511 ± 79}& a & 61.9 ± 4.7  & 15.481 ± 0.036\\
    WISE-1738  & WISEPA J173835.53$+$273258.9& 3 & 450 \loicadd{± 88}& a & 130.9 ± 2.1 & 14.519 ± 0.017\\
    WISE-2056  & WISEPC J205628.90$+$145953.3& 3 & \loicadd{$481_{-20}^{+26}$}& c & 140.8 ± 2.0 & 13.925 ± 0.016\\
    WISE-2220  & WISE J222055.31$-$362817.4  & 2 & \loicadd{452 ± 88}& a & 95.5 ± 2.1  & 14.807 ±	0.024\\
    WISE-2354  & WISEA J235402.79$+$024014.1 & 8 & \loicadd{$347_{-11}^{+13}$}& c & 130.6 ± 3.3 & 15.018 ± 0.030\\
    \multicolumn{7}{c}{\textbf{NIRISS F480M Imaging (Program ID 1230)}} \\
    UGPS-0722  & UGPS J072227.51$-$054031.2 & 10 & \loicadd{569 ± 45}& a & 242.8 ± 2.4 & 12.198 ± 0.008\\
    WISE-0855  & WISE J085510.83$-$071442.5 & 11 & \loicadd{250 ± 50}& a & 439.0 ± 2.4 & 13.820 ± 0.029\\
    \multicolumn{7}{c}{\textbf{NIRISS F480M Imaging (Program ID 1189)}} \\
    WISE-1828  & WISEPA J182831.08$+$265037.8&3 & \loicadd{406 ± 88}& a & 100.3 ± 2.0 & 14.393 ± 0.016\\

    \hline
 \end{tabular}
 \tablecomments{
 Discovery references are --
    1: \citet{Pinfield_2014a}; 
    2: \citet{Kirkpatrick_2012}; 
    3: \citet{Cushing_2011}; 
    4: \citet{Kirkpatrick_2013}; 
    5: \citet{Griffith_2012}; 
    6: \citet{2020_Bardalez};
    7: \citet{Tinney_2014}; 
    8: \citet{Schneider_2015}; 
    9: \citet{marocco_2020};
    10: \citet{Lucas_2010}; 
    11: \citet{Luhman_2014}.\\
 $T_{\rm eff}$ references are --
    a: \citet{kirkpatrick_2024}; 
    b: \citet{Kirkpatrick_2021}; 
    c: \citet{Beiler_2024}. 
    \\
    Parallaxes are from \citet{Kirkpatrick_2021} \loicadd{except for WISE-1446 where the parallax is from \citet{Beiler_2024}. \\
    $^{\dagger}$ Updated astrometry for WISE-0830 is given in Table~\ref{tab:W0830newmodel}.}\\
    The W2 magnitudes are proper-motion corrected values ({\tt w2mpro\_pm}) from the CatWISE2020 Catalog \citep{Marocco_2021}, 
    curated by IRSA.
    }
\end{table*}

The target list of Y dwarfs was selected from the literature as available in 2020. We selected spectroscopically confirmed brown dwarfs as listed in \citet[Table 11]{kirkpatrick_2019} with $T_{\rm eff} < 500$~K within 15~pc of the Sun. For the NIRCam program (PID 2473), brown dwarfs calculated to be too close to known background objects during Cycle 1 were removed from the sample as were known binary brown dwarfs. Two targets in that sample were observed for PI Albert as part of GTO programs (PIDs 1189 and 1230) using NIRISS in imaging mode which included two additional targets in the 15-pc Y-dwarfs sample (WISE\,J085510.83$-$071442.5 and WISEPA\,J182831.08$+$265037.8) as well as an additional PSF calibration target, UGPS\,J072227.51$-$054031.2, a nearby T9 brown dwarf. \loicadd{The full sample is listed in Table 1 with discovery references, effective temperature values, parallax measurements, and CatWISE 2020 W2 magnitudes.} Hereafter, the targets will be written as WISE-hhmm, e.g. WISE-0855.

\subsection{{\it JWST} NIRCam Imaging}

NIRCam enables dual band imaging in the blue ($\lambda\leq2.4\,\mu$m) and red ($\lambda\geq2.5\,\mu$m) channels simultaneously \citep{Rieke_2023}. For the blue channel, we selected the F150W (1.5$\mu$m) wide band filter ($\Delta\lambda/\lambda=10$\%) because most Y dwarfs had measurements in the ground-based $H$-band filter which is centered near 1.6\,$\mu$m (c.f. \cite{kirkpatrick_2019,Kirkpatrick_2021}). For the red channel we selected a filter at a wavelength where the Y dwarfs emit significant flux, the F480M (4.8$\mu$m) medium band filter ($\Delta\lambda/\lambda=5$\%) \citep{Schneider_2015,Morley_2014}.

We observed the sample in 5 groups of 4 targets and specified that observations within a group were to be executed within 3 days. This strategy was designed to minimize any wavefront drift and, for each target, to be able to use the 3 other targets as point source function (PSF) reference stars (assuming they were not binaries). 

Integration times range, for a given target, between 30 minutes and 2.5 hours which provided a signal-to-noise ratio (SNR) $>$ 1000 in F480M. Achieving such a high SNR (or roughly 10$^7$ photons) at 4.8~$\mu$m was crucial to be able to use the kernel phase interferometry analysis technique for our binary search. The SNRs in F150W were much lower (10 to 100) because of the extremely red $H-$W2 colors of $\geq6$~magnitudes, for this sample of Y dwarfs. Note that this high SNR is considered separately from the {\em absolute} flux calibration uncertainty for NIRCam currently listed at $0.9\%$ (F480M) and $0.4\%$ (F150W) in the respective {\em photom} reference files found on the {\em JWST Calibration Reference Data System}\footnote{https://jwst-crds.stsci.edu/} (CRDS).

Each target was observed over the $2\farcm2\times 2\farcm2$ field of view of module B in FULL readout mode with dithered exposures. The first 12 targets used a 5-point subpixel dither type ({\tt SMALL-GRID-DITHER}) that was found to be suboptimal for bad pixel correction. It was switched to a larger (2\farcs9-8\farcs9) 5-point primary dither strategy ({\tt INTRAMODULEBOX}) enabling better bad pixel interpolation for the remaining 8 targets. We attempted to center the target near pixels 1200, 700 on the long wave detector (F480M band) corresponding to pixels near 400,1400 in the short wave detector NIRCB2 (F150W band) by using a target offset of $\Delta\alpha,\Delta\delta = +22\farcs0,-8\farcs0$. Errors in the position epoch entered in the \emph{Astronomer Proposal Tool} caused significant deviations, in one case resulting in having the target off the F150W channel for 2 dithers. Table~\ref{tab:NewData} lists the observations for targets observed as part of Program 2473.

\subsection{{\it JWST} NIRISS Imaging}

The target list was supplemented with three additional brown dwarfs, UGPS-0722, WISE-0855 and  WISE-1828, observed with NIRISS in imaging mode using filters F480M + CLEARP to search for companions (see Table~\ref{tab:NewData}). One of the sources, UGPS-J0722, is a late T dwarf and not a Y dwarf, and is warmer than the rest of the sample, with $T_{\rm eff} \approx 540$~K \citep[e.g.][]{Leggett_2021}.

The WISE-0855 observation was accompanied by a contemporaneous calibrator star observation of UGPS-0722, intended to serve as a PSF reference for both WISE-0855 and WISE-1828. During the first epoch of observations of WISE-0855 and UGPS-0722 the Fine Guidance Sensor lost the guide star lock and repeat observations of both targets were performed. A large SNR ($>1000$) was achieved for all three targets. Observations are un-dithered (staring) and centered on a bad-pixel-free region of the NIRISS detector. All observations with NIRISS were obtained in FULL subarray mode, except UGPS-0722, which was observed in SUB80 (80$\times$80 pixels) to prevent saturation.

\bigskip
\section{Data Reduction and Photometry}

\begin{table*}[!ht]
\centering
\caption{F150W and F480M Photometry of our Y dwarf Sample \label{tab:NewData}}
\begin{tabular}{lcrll}
\hline
Short Name & Observing Date & Int. Time & F150W & F480M \\
           & Mid Exposure (UT) & (minutes) & (Vega Mag) & (Vega Mag) \\
\hline \hline
WISE-0304 &  2022-09-23T21:58:02 &  89.5 & 21.872 $\pm$ 0.010 & 15.428 $\pm$ 0.005\\
WISE-0336AB& 2022-09-22T13:13:10 &  40.3 & 22.038  $\pm$ 0.012  & 14.538  $\pm$ 0.005 \\
WISE-0336A$^a$ & 2022-09-22T13:13:10 &  40.3 & 22.114  $\pm$ 0.016  & 14.727$^{+0.054}_{-0.023}$ \\
WISE-0336B$^a$ & 2022-09-22T13:13:10 &  40.3 & 24.938$^{+0.175}_{-0.105}$    & 16.534$^{+0.120}_{-0.252}$   \\
WISE-0359 &  2022-09-23T17:51:38 &  89.5 & 22.281 $\pm$ 0.010 & 15.268 $\pm$ 0.005\\
WISE-0410 &  2022-09-23T23:43:56 &  26.8 & 20.167 $\pm$ 0.010 & 13.999  $\pm$ 0.005\\
WISE-0535 &  2022-06-27T22:27:30 &  62.6 & 22.866 $\pm$ 0.013 & 14.795  $\pm$ 0.005\\
WISE-0647 &  2022-11-11T16:38:29 &  58.2 & 23.332 $\pm$ 0.013 & 14.919 $\pm$ 0.005\\
WISE-0713 &  2022-11-12T08:42:33 &  31.3 & 20.445 $\pm$ 0.010 & 14.183 $\pm$ 0.005\\
UGPS-0722$^b$ &  2023-04-17T22:33:44 &  18.9 & 17.26  $\pm$ 0.03  & 12.016  $\pm$ 0.011 \\
WISE-0734 &  2022-06-28T00:07:35 &  76.1 & 21.706 $\pm$ 0.011 & 15.162  $\pm$ 0.004\\
WISE-0825 &  2022-11-12T10:10:45 &  40.3 & 22.831 $\pm$ 0.014 & 14.418 $\pm$ 0.005\\
WISE-0830 &  2022-11-12T22:49:59 & 120.8 & 25.261 $\pm$ 0.026 & 15.644 $\pm$ 0.005\\
WISE-0855$^b$ &  2022-11-19T09:25:29 &  36.5 & 23.98  $\pm$ 0.10  & 13.679  $\pm$ 0.011 \\
WISE-1141 &  2023-06-29T22:27:31 &  40.3 & 20.580 $\pm$ 0.010 & 14.582 $\pm$ 0.005\\
WISE-1206 &  2023-02-27T21:31:43 &  71.6 & 21.227 $\pm$ 0.010 & 15.112 $\pm$ 0.005\\
WISE-1405 &  2023-03-02T12:43:40 &  25.1 & 21.756 $\pm$ 0.012 & 13.930 $\pm$ 0.005\\
WISE-1446 &  2023-03-02T09:15:03 & 143.2 & 23.672 $\pm$ 0.013 & 15.694  $\pm$ 0.005\\
WISE-1541 &  2023-03-02T11:18:04 &  28.6 & 22.004 $\pm$ 0.011 & 14.018 $\pm$ 0.005\\
WISE-1639 &  2023-06-28T21:47:19 &  13.4 & 21.316 $\pm$ 0.014 & 13.487 $\pm$ 0.005\\
WISE-1738 &  2022-06-29T17:21:58 &  31.3 & 20.365 $\pm$ 0.010 & 14.464 $\pm$ 0.005\\
WISE-1828$^b$ &  2022-07-28T17:52:53 &  35.8 & 23.07   $\pm$ 0.10  & 14.242 $\pm$ 0.011 \\
WISE-2056 &  2023-06-28T09:03:58 &  22.4 & 20.057 $\pm$ 0.001 & 13.762  $\pm$ 0.005\\
WISE-2220 &  2023-07-01T21:31:28 &  49.2 & 21.161 $\pm$ 0.011 & 14.745 $\pm$ 0.005\\
WISE-2354 &  2022-06-29T19:09:18 &  58.2 & 23.098 $\pm$ 0.013 & 14.930 $\pm$ 0.005\\
\hline
    \end{tabular} 
        \tablecomments{
        $^a$ The individual F150W and F480M magnitudes for the WISE-0336 system were determined from the aperture photometry for the unresolved system and the flux ratios for each component measured by \citet{Calissendorff_2023}.\\
        $^b$ The F150W magnitudes for UGPS-0722, WISE-0855 and WISE-1828 were synthesized from spectroscopic  observations by \citet{Lucas_2010, Luhman_2024, Cushing_2021}. Uncertainties for these values were estimated to be equal to the spectral flux calibration uncertainty.} 
\end{table*}

\subsection{NIRCam}

We downloaded the stacked images and associated catalogues constructed from the 5 dithered exposures ({\em i2d.fits}) processed by the Data Management System (DMS) \citep{bushouse.2022}, version 1.9.6. The photometry catalogues produced by the default DMS level 3 pipeline contained anomalous magnitude entries for a few targets which prompted us to revisit this step by performing our own aperture photometry using the photutils package version 1.8.0 \citep{photutils_1.8.0}. For each filter (F480M/F150W), we first determined the centroid of the PSF by fitting a quadratic function in the 3$\times$3 pixels around the pixel with peak intensity, then estimated the local sky level from statistics in an annulus with radii of 4.92/6.082 and 7.083/9.496 pixels around the PSF center and subtracted this level. Next we performed aperture photometry by summing pixel flux within a radius of 3.757/3.199 pixels, defined in the CRDS reference file, {\tt jwst\_nircam\_apcorr\_0004.fits}, as the radius encompassing 70\% of the PSF encircled energy. Finally, we applied an aperture correction by multiplying the extracted flux by 1.4863/1.4485. This yielded flux measurements, F, expressed in megajansky/steradian that we converted to Vega magnitudes by applying the $mag_{AB} - mag_{\tt Vega} = -3.85$/$-2.15$\, mag offset (reference: {\tt jwst\_nircam\_abstovegaoffset\_0002.fits}) from the AB magnitude:
\begin{equation}
mag_{AB} = -2.5~log_{10}\left(\frac{F [MJy/sr] \times T \times A [sr/pixel]}{3631 [Jy] \times10^{-6} [MJy/Jy]}\right)
\end{equation}
where T is the throughput calibration, PHOTMJSR $= 1.456\pm0.006$/$2.338\pm0.021$, A is the steradian to pixel area conversion, PIXAR\_SR $= 9.332\times10^{-14}$/$2.287\times10^{-14}$, found in the {\tt jwst\_nircam\_photom\_0153/0150.fits} reference file.

The measurement uncertainty considers the readout noise and photon noise given as a pixel map by the DMS pipeline (the ERR extension of the cal.fits) but also includes the uncertainty in subtracting the background level. These combine to very small statistical errors of order 0.08\% (F480M) and 0.2-2\% (F150W). The absolute flux calibration uncertainty is added in quadrature and, with a precision of 0.4\% in the F480M, dominates the error budget for the photometric measurements reported in Table~\ref{tab:NewData}. For the F150W filter, the absolute calibration uncertainty, 0.9\%, is comparable to the statistical noise.

\subsection{Photometry of the binary WISE~0336A+B}

We present updated photometry for both components of the known binary WISE-0336A/B from that reported in \cite{Calissendorff_2023}. We treated the A+B components as unresolved to obtain the aperture photometry of the system in exactly the same manner as for the other NIRCam targets. We then used the binary contrast published in \cite{Calissendorff_2023} of $\Delta F480M = F480M_B - F480M_A = 1.81^{+0.14}_{-0.31}$ and $\Delta F150W = F150W_B - F150W_A = 2.82^{+0.19}_{-0.11}$ to calculate the magnitude of each component separately. The uncertainties were propagated using a Monte Carlo simulation of $5\times10^5$ realizations. Results are resilient to aperture size selection.

\subsection{NIRISS}

We reprocessed the data with the DMS version 1.12.5 enabling the {\tt charge\_migration} step, up to the end of level 2 which produced {\em cal.fits} images. Then we ran a custom 1/f correction on the 3 data sets observed in FULL mode (both WISE-0855 observations and WISE-1828 observation) while using directly the {\em cal.fits} for UGPS-0722 because the 1/f imprint is less pronounced on the SUB80 images. Using the custom 1/f corrected data changed the photometry by less than 1\% compared to using the {\em cal.fits} directly.

To extract photometry of NIRISS F480M images, we adopted the same procedure as performed on the NIRCam data sets and used by the DMS pipeline. We used the prescribed aperture and sky annulus radii of 4.68, 7.80 and 11.74 pixels, respectively while the aperture correction is 1.458 (70\% encircled energy), found in jwst\_niriss\_apcorr\_0008.fits. The AB to Vega magnitude offset is $mag_{AB} - mag_{\tt Vega} = -3.4268$, the PHOTMJSR $= 1.220\pm0.013$ and PIXAR\_SR $=1.009\times10^{-13}$ (jwst\_niriss\_abstovegaoffset\_0003.fits, jwst\_niriss\_photom\_0043.fits). The photometry has very small statistical errors of order 0.07\%. The absolute flux calibration uncertainty is added in quadrature and, with a precision of 1.04\%, dominates the error budget for the \loicadd{the F480M} photometric measurements reported in Table~\ref{tab:NewData}.

\begin{table*}[!ht]
    \centering
        \caption{Astrometry for the NIRCam Sample}
    \label{tab:astrometry}
    \begin{tabular}{lrrrrrrr}
    \hline
    Short & MJD & $\alpha_{2000}$ & $\delta_{2000}$ & No. of Gaia & \multicolumn{3}{c}{Offset from K21 Model} \\
    Name & (days) & (deg(mas)) & (deg(mas)) & Anchors & $\alpha$ (mas) & $\delta$ (mas) & $\sigma^a$ \\
    \hline
    \hline

WISE-0304 & 59845.91531 &   46.2048671( 20.0) &  -27.0839387( 15.8) &    4 &  $       13.9\pm       21.7$ &  $      -37.3\pm       19.7$ &   2.0 \\ 
WISE-0336 & 59844.55082 &   54.0202647( 16.4) &   -1.7351151( 15.2) &    6 &  $      -18.4\pm       17.9$ &  $      -12.3\pm       16.7$ &   1.3 \\ 
WISE-0359 & 59845.74420 &   59.8912978( 15.0) &  -54.0343645( 15.2) &    3 &  $        2.6\pm       15.7$ &  $       -3.8\pm       16.4$ &   0.3 \\ 
WISE-0410 & 59845.98885 &   62.5983013( 15.8) &   15.0390435( 19.0) &    4 &  $        3.1\pm       17.1$ &  $        1.3\pm       20.0$ &   0.2 \\ 
WISE-0535 & 59757.93577 &   83.8183968( 18.2) &  -75.0066568( 16.7) &  110 &  $        4.2\pm       19.1$ &  $       25.5\pm       17.9$ &   1.4 \\ 
WISE-0647 & 59894.69340 &  101.8468599( 15.4) &  -62.5418782( 17.4) &   13 &  $       10.5\pm       15.9$ &  $       -5.7\pm       17.9$ &   0.7 \\ 
WISE-0713 & 59895.36289 &  108.3454431( 16.6) &  -29.2992193( 17.4) &   98 &  $       -4.6\pm       18.0$ &  $      -17.7\pm       18.6$ &   1.0 \\ 
WISE-0734 & 59758.00527 &  113.6771949( 16.6) &  -71.9624680( 16.9) &   14 &  $        0.7\pm       16.9$ &  $       -6.8\pm       17.4$ &   0.4 \\ 
WISE-0825 & 59895.42414 &  126.2804168( 15.5) &   28.0959557( 15.9) &    7 &  $        5.5\pm       17.0$ &  $       -4.5\pm       17.2$ &   0.4 \\ 
WISE-0830 & 59895.95138 &  127.5492249( 29.2) &   28.6142199( 23.1) &    4 &  $      182.9\pm      309.9^b$ &  $      112.1\pm      198.7^b$ &   0.8 \\ 
WISE-1141 & 60124.93578 &  175.4823320( 18.6) &  -33.4434263( 20.2) &   15 &  $       32.7\pm       22.2$ &  $      -12.7\pm       23.5$ &   1.6 \\ 
WISE-1206 & 60002.89703 &  181.4985345( 16.0) &   84.0186397( 23.0) &    3 &  $        2.0\pm       17.7$ &  $       16.9\pm       24.0$ &   0.7 \\ 
WISE-1405 & 60005.53033 &  211.3120210( 16.9) &   55.5733955( 15.1) &    4 &  $       -1.0\pm       19.3$ &  $       13.2\pm       16.8$ &   0.8 \\ 
WISE-1446 & 60005.38545 &  221.5254686( 20.8) &  -23.2903901( 15.6) &   14 &  $      195.8\pm      323.1$ &  $      113.0\pm      168.5$ &   0.9 \\ 
 &  &  &   &    &  $       52.4\pm       85.4^c$ &  $       60.9\pm       80.8^c$ &   1.0$^b$ \\ 
WISE-1541 & 60005.47089 &  235.4616164( 17.1) &  -22.8407867( 17.1) &   31 &  $        3.7\pm       18.6$ &  $       29.7\pm       18.6$ &   1.6 \\ 
WISE-1639 & 60123.90787 &  249.9258850( 16.3) &  -68.8055121( 15.8) &  106 &  $       -4.4\pm       18.4$ &  $      -24.3\pm       18.1$ &   1.4 \\ 
WISE-1738 & 59759.72360 &  264.6493275( 16.4) &   27.5485376( 15.6) &   19 &  $       -7.9\pm       17.4$ &  $        8.9\pm       16.8$ &   0.7 \\ 
WISE-2056 & 60123.37776 &  314.1235676( 16.5) &   15.0000852( 16.2) &   44 &  $       19.1\pm       17.8$ &  $        3.0\pm       17.6$ &   1.1 \\ 
WISE-2220 & 60126.89686 &  335.2318522( 21.2) &  -36.4718910( 17.5) &    6 &  $       -0.6\pm       22.7$ &  $       -3.0\pm       18.9$ &   0.2 \\
WISE-2354 & 59759.79813 &  358.5132467( 20.6) &    2.6693245( 16.3) &    3 &  $      -46.6\pm       24.0$ &  $      -39.8\pm       20.3$ &   2.8 \\ 

    \hline
    \end{tabular}
 \tablecomments{$^a$ The last column (Offset from K21 Model - $\sigma$) compares the measured and modelled positions and expresses the offset in units of the combined model + measurement uncertainties. 
 $^b$ For WISE-0830 the model was updated; results are given in Table~\ref{tab:W0830newmodel}.
 $^c$ For WISE-1446, comparison with the model of Marocco et al. (in prep.) shows good agreement and improved precision.}
\end{table*}

\begin{table}[]
    \centering
        \caption{Updated Spitzer + JWST astrometry model for WISEA J083011.95$+$283716.0.}
    \begin{tabular}{rl|c}
    \hline
    \multicolumn{2}{c}{Model parameter} & Value ($\chi^2_\nu = 0.749$, 18 data points) \\
    \hline \hline
    Epoch & (t$_0$) & 58850 (MJD) \\
    R.A. at t$_0$ & ($\alpha_{\rm ICRS}$) & $127.549366\degree \pm 6.8$\,mas \\
    Decl. at t$_0$ & ($\delta_{\rm ICRS}$) & $28.615822\degree \pm 5.6$\,mas \\
    Parallax & ($\varpi$) & $99.2\pm6.5$\, mas \\
    P.M. R.A. & ($\mu_{\rm R. A.}$) & $-190.2\pm8.4$\, mas yr$^{-1}$ \\
    P.M. Decl. & ($\mu_{\rm Decl.}$) & $-2011.0\pm6.9$\, mas yr$^{-1}$ \\
    \hline
    \end{tabular}
    \label{tab:W0830newmodel}
\end{table}

\loicadd{In order to compare the colors of these three NIRISS sources to the rest of our sample, we  synthesized F150W photometry from available near-infrared spectra} \citep{Lucas_2010, Luhman_2014, Cushing_2021}.  
\loicadd{The {\em JWST} user documentation supplies 
filter transmissions which
include instrument system throughput for each camera; we used the F150W NIRCam transmission for consistency with the other measurements. For reference, we explored the differences between the NIRCam and NIRISS photometric systems by synthesizing F150W and F480M magnitudes for each camera for five Y dwarfs with {\it JWST} spectra across the bandpasses \citep{Beiler_2024}.  The five Y dwarfs had a range in $T_{\rm eff}$ of 350~K to 500~K,  and the differences in the photometry were small, with no trend seen with $T_{\rm eff}$. We found an average difference for NIRCam $-$ NIRISS of $\Delta F150W = -0.040 \pm 0.012$ and $\Delta F480M = -0.010 \pm 0.001$ magnitudes. Any systematic difference due to the different F480M throughput for the two cameras are therefore similar in size to the absolute flux calibration uncertainty.  
}

\bigskip
\section{Astrometry --- Proper Motion and Parallax}

The most recent parallax measurements place the brown dwarfs of our sample at a distance of 16~pc, or closer, to the Sun. All have significant proper motion, mostly derived from {\it Spitzer} observations obtained between 2010 and 2019 \citep{Kirkpatrick_2021}. Our {\it JWST} observations obtained in 2022 and 2023 can therefore test the published astrometry by extrapolating the proper motion and parallax model over a 3 - 4 year time frame.


To propagate the motion of our targets proper motion and parallax, we use the SkyCoord module of Astropy and convert celestial coordinates to the Geocentric Celestial Reference System (GCRS). To eliminate the annual aberration which is the dominant motion, we measure angular positions relative to a virtual distant reference point at the celestial position of each target. We neglect the parallax due to {\it JWST} being at the Lagrange point 2 which introduces parallax errors of less than 1\%, or typically $\sim1$~mas at the distance of our targets.

The initial position measurements of all sources including the brown dwarfs ({\tt RA\_ICRS, DE\_ICRS}) come from the catalogue produced by the default DMS level 3 product on the stacked frame. \loicadd{We downloaded the most recent version of the catalogues found on MAST, based on the calibration software version 1.15.1.} In each field, the astrometry is then anchored on available Gaia DR2 sources (from 3 to 100), taking proper motion and parallax into account.  \loicadd{Experimentations with 4 targets having more than 40 Gaia stars in the NIRCam field of view show that, after applying an offset in right ascension and declination, residuals appear unstructured with no obvious rotation or scale pattern. Also, for about half our targets, there is insufficient Gaia-measured stars ($\leq 7$) to confidently constrain the 4-parameter affine transformation (offsets, rotation, scale). Therefore, for the purpose of this astrometry check, anchoring only considers right ascension and declination offsets.} Anchoring uncertainties range between \loicadd{15 and 29 mas} per axis,  (See Table~\ref{tab:astrometry}). 


As our baseline model for the brown dwarf astrometry, we adopt the \citet[hereafter K21 in the text and figures]{Kirkpatrick_2021} proper motion and parallax\loicadd{. In the case of WISE-1446, we adopt the more precise parallax and proper motion measurements of Marocco et al. (in prep.).} \loicadd{These models include an epoch of reference, t$_0$, and corresponding celestial positions, $\alpha_{\rm ICRS}$, $\delta_{\rm ICRS}$ at t$_0$.} \loicadd{To check predictions of the model, we propagate the target position to the epoch of our {\it JWST} observation} and compare it with our measured position. The propagated position uncertainties are obtained by performing a Monte Carlo simulation using 1000 trials. \loicadd{The uncertainty of each model parameter (parallax, proper motion, t$_0$, $\alpha_{\rm ICRS}$ and $\delta_{\rm ICRS}$) is assumed to be a gaussian distribution centered on the parameter value} \loicadd{For all but one brown dwarf, WISE-2354, we confirm that the position of the target in the JWST images agrees with predicted positions within 2$\sigma$. The disagreement for WISE-2354 is at the $2.8\sigma$ level. That is too small to impact future JWST follow ups as, in absolute term, the error is well under 100\,mas or roughly one NIRCam long-wave pixel. }

\loicadd{In general, the propagated model uncertainties at the epoch of our {\it JWST} observation (2022) are very small (4 to 12 mas per axis). However, the K21 model for two brown dwarfs, WISE-0830 and WISE-1446, 
has a large total uncertainty of $\geq$ 350\,mas.  In the case of WISE-0830, we decided to update the model because even a single JWST measurement offered the opportunity to improve its precision,  especially as the JWST measurement happened to sample a parallax peak. We ran the existing tool described in \citet{kirkpatrick_2019, Kirkpatrick_2021} by combining this measurement to the existing suite of Spitzer and WISE measurements already available. The  {\it JWST} spacecraft barycentric (XYZ) position was considered in our calculations. Results are presented in Table~\ref{tab:W0830newmodel}. The new parallax places WISE-0830 at a distance of $10.1_{-0.6}^{+0.7}$\,pc, about 10\% closer than previously thought ($11.0_{-1.5}^{+2.0}$\,pc).  For WISE-1446, we were able to test a recent and more precise astrometry model (Marocco et al. in prep.) which shows a better agreement with our observation and has an improved precision of $\leq$ 100\,mas. }

\bigskip
\section{Brown Dwarf Colors}

\begin{figure*}[!ht]
    \centering  
    \vskip -0.5in
\includegraphics[angle=0,width=0.85\textwidth]{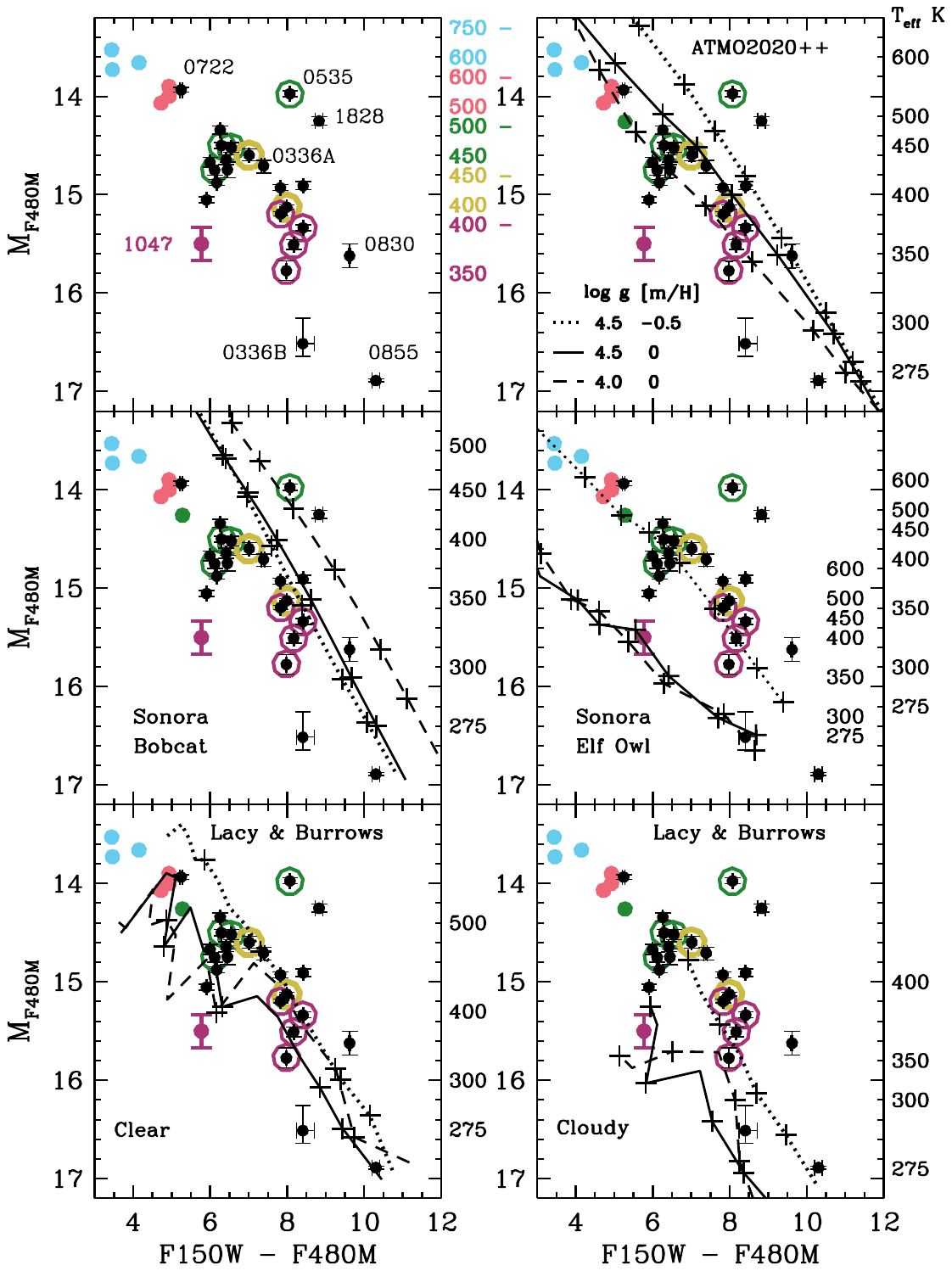}
    \vskip -0.5in
    \caption{F480M vs. F150W$-$F480M color-magnitude diagrams. Black circles represent new measurements presented here. Colored dots are synthesized colors from \citet{Beiler_2024} where the symbol color indicates $T_{\rm eff}$, determined by Beiler et al. from luminosity measurements. Colored rings similarly indicate $T_{\rm eff}$; the symbol color for a given $T_{\rm eff}$ range is 
    \loicadd{indicated along the right axis of the}
     top left panel. 
    \loicadd{The warmest and coldest objects in our sample, the WISE-0336 binary system, and outliers, } 
    are identified in the top left panel. The other panels compare the data to various model sequences; surface gravity and metallicity properties are indicated by line type as shown in the legend in the top right panel. Sequences in the top right panel are from disequilibrium chemistry and adiabat-adjusted ATMO2020++ models \citep{Leggett_2021, Meisner_2023}. The middle panels show Sonora sequences, equilibrium-chemistry Bobcat on the left  \citep{Marley_2021}, and on the right disequilibrium chemistry Elf Owl \citep{Mukherjee_2024}. The bottom panels show disequilibrium chemistry models from  \citet{Lacy_2023} with and without water clouds. Model
    $T_{\rm eff}$ values are indicated along the right axes and plus symbols along the model sequences indicate colors at those $T_{\rm eff}$.  \loicadd{For the Elf Owl panel the outer axis $T_{\rm eff}$ values refer to the metal-poor sequence and the inner values to the solar metallicity sequences; for the other models the $M_{480M}$ values are more similar across the metallicity range.}
    }
    \label{fig:CMD}
\end{figure*}

Figure~\ref{fig:CMD} shows a color magnitude diagram with the new data presented here as black filled circles. Our homogeneous {\it JWST} photometry confirms the large intrinsic scatter ($\sim 1$~mag) previously seen for Y dwarfs in color-magnitude diagrams \citep[e.g.][]{Kirkpatrick_2021, Leggett_2021}, which excludes measurement noise or filter transformations as an explanation. 

Other filled circles in Figure~\ref{fig:CMD}  represent \loicadd{F150W and F480M colors synthesized from spectra by \citet[][their Tables 7 and 9]{Beiler_2024}, using NIRCam bandpasses.}
The symbol color indicates the $T_{\rm eff}$ value determined by \citet{Beiler_2024} \loicadd{via bolometric luminosities determined from} empirical SEDs built with MIRI and NIRSpec spectra. Colored rings around black points  indicate the \citet{Beiler_2024} $T_{\rm eff}$ values for those sources. The  \loicadd{empirical \citet{Beiler_2024} $T_{\rm eff}$ range for each symbol color is indicated along the right axis of the top left panel.}

\loicadd{\subsection{Comparison to Models}}

\loicadd{We compare the photometry to synthetic photometry calculated by four cool brown dwarf models, each based on different atmosphere prescriptions.}
As a baseline model, the Sonora Bobcat models assume equilibrium chemistry \citep{Marley_2021}. But, given the evidence for out-of-equilibrium chemistry in the atmosphere of ultracool dwarfs, the Sonora Elf Owl models include disequilibrium chemistry \citep{Mukherjee_2024}, the ATMO2020++ models use disequilibrium chemistry and introduce adjustments to the adiabat profile \citep{Leggett_2023}, while \cite{Lacy_2023} introduce water clouds in their disequilibrium models because the atmospheres of the coldest Y dwarfs, those with $T_{\rm eff} \lesssim 400$~K, are cold enough for water to condense \citep[e.g.][]{Morley_2014}. \loicadd{We note that the Sonora Cholla \citep{Karalidi_2021} and Diamondback \citep{Morley_2024} models do not include atmospheres as cold as those of the Y dwarfs.}

\loicadd{All} model sequences  in Figure~\ref{fig:CMD}  are cloud-free except for the \citet{Lacy_2023} sequence shown \loicadd{in the bottom right panel.}. Solar and sub-solar metallicities are included. Surface gravities of log $g = 4.0$ cm~s$^{-2}$ and 4.5 cm~s$^{-2}$ are shown. From evolutionary models we estimate that field dwarfs with ages of 2 to 4 Gyr have log $g \approx 4.5$ cm~s$^{-2}$ for $325 \lesssim T_{\rm eff}$~K $\lesssim 550$ and log $g \approx 4.0$ cm~s$^{-2}$ for $200 \lesssim T_{\rm eff}$~K $\lesssim 325$ \citep{Marley_2021}. \loicadd{Model sequences' line types indicate metallicity and gravity, as shown in the legend in the top right panel.}

 \loicadd{Figure~\ref{fig:CMD} indicates that} the ATMO2020++  \citep{Phillips_2020,Leggett_2021, Meisner_2023, Leggett_2024a} \loicadd{and the \citet{Lacy_2023} models}  reproduce the observed \loicadd{colors} better than the Sonora Bobcat or Elf Owl models \citep{Marley_2021, Mukherjee_2024}.  
The $T_{\rm eff}$ values from the
ATMO2020++ \loicadd{and the \citet{Lacy_2023} models}
are also in good agreement with the values measured by \citet{Beiler_2024} from the observed luminosities,
\loicadd{while the Bobcat and Elf Owl models are not.} 

\loicadd{For brown dwarfs warmer than 350~K, the Sonora Bobcat F150W $-$ F480M colors appear to be too red, while the Elf Owl colors appear to be too blue (assuming that this local sample of brown dwarfs is not unusually metal-poor). The likely cause of the red Bobcat color is the exclusion of disequilibrium chemistry, which increases the abundance of CO and CO$_2$ \citep[e.g.][]{Zahnle_2014} which absorb light at  $\lambda \approx 4.7~\mu$m and 4.2~$\mu$m respectively  \citep[e.g.][their Figure 5]{Tu_2024}, thereby decreasing the F480M flux. The likely cause of the blue Elf Owl color is an excess of near-infrared flux at F150W; it seems that standard radiative-convective atmospheres for cold brown dwarfs produce an excess of energy in the near-infrared \citep[e.g.][]{Leggett_2017,
Leggett_2021,Karalidi_2021}, which is  the issue that the ATMO2020++ models address empirically by changing the pressure-temperature profile such that the deeper regions, where the near-infrared energy emerges, is colder. Although empirical, the adjustment is physically motivated by thermal changes due to rapid rotation and/or chemical changes in the deep layers due to condensation \citep[e.g.][]{Leggett_2021, Leggett_2024a}.
}

The ATMO2020++ sequence does appear to show a systematic offset in color.   This may be partly due to the definition of the F150W bandpass. For cold brown dwarfs the signal through this filter comes only from the narrow $H$-band flux peak at the extreme red of the bandpass. Hence the color is sensitive to the definition of the red cut-off of the filter.

The ATMO2020++ \loicadd{and the \citet{Lacy_2023}} models calculate that the F480M absolute magnitude is a good indicator of $T_{\rm eff}$ for 275~K to 600~K brown dwarfs because of its \loicadd{relatively} small dependence on metallicity and gravity ($\Delta M_{\tt F480M} \sim$ \loicadd{0.3}~mag for $\Delta$ log~$g$ or $\Delta $[m/H] $\sim 0.5$~dex) --- compare the $y$-axis locations of the plus symbols  in Figure~\ref{fig:CMD}. On the other hand, these models calculate that the F150W \loicadd{$-$ F480M color} is sensitive to 
\loicadd{both gravity and metallicity}

\loicadd{($\Delta ({\tt F150W - F480M}) \sim$  \loicadd{1.0}~mag for $\Delta$ log~$g$ or $\Delta $[m/H] $\sim 0.5$~dex)}
 --- compare the $x$-axis locations of the plus symbols in Figure~\ref{fig:CMD}.

\loicadd{\subsection{Superluminous or Very Red Brown Dwarfs}}

\loicadd{WISE-0535 and  WISE-1828 appear to be superluminous and/or redder than the other dwarfs in Figure~\ref{fig:CMD}. WISE-0830 may be a colder example of this group. We discuss each of these brown dwarfs individually  below, in RA order.}

A preliminary analysis of the spectral distribution of WISE-0535 using ATMO2020++ synthetic spectra by \citet{Leggett_2024b} indicates that its super-luminosity is due to multiplicity. We address this further in our companion paper Vandal et al. in prep.

This work and Matuszewska et al. (in prep.) present the first near-infrared detection of the cold brown dwarf WISE-0830. Its location in Figure~\ref{fig:CMD} suggests that it is a typical field dwarf with \loicadd{$T_{\rm eff} \approx 350~K$}. However 
WISE-0830 may be superluminous, \loicadd{depending on the, currently poorly-defined, color-magnitude trend for the coldest objects. Superluminosity}
suggests lower metallicity, higher gravity (older age), or \loicadd{multiplicity}.  WISE-0830 has the highest tangential velocity in our sample \citep[108~km~s$^{-1}$, ][]{2020_Bardalez} possibly supporting an older age and lower metallicity for this Y dwarf, compared to typical field dwarfs.

The extreme colors of WISE-1828 may indicate that it is multiple, has a high gravity, or is metal-poor. Recent studies using {\em JWST} imaging data find no evidence of a companion at separations $> 0.5$~au \citep{Furio_2023}.  Recent spectral analyses of {\em JWST} NIRSpec and MIRI data by  \cite{lew.2024} and \citep{Barrado_2023} find that the metallicity is approximately solar, and that the luminosity is consistent with evolutionary models if the system is a tight binary --- an order-of-magnitude estimate by \citet{lew.2024}, based on the radial velocity, suggests a separation of 20 Jupiter radii.  \citet{Leggett_2024a}
find a good fit to the {\it JWST} \loicadd{NIRSpec and MIRI} data with ATMO2020++ solar metallicity 
models if the system is an equal-mass binary with unusually high gravity i.e. a relatively massive and old system.

\loicadd{\subsection{A Blue Candidate Very Young  Jupiter-Mass Brown Dwarf}}

CWISEP J104756.81+545741.6 (hereafter WISE-1047, identified as ``1047'' in Figure~\ref{fig:CMD}) is
not in our imaging sample, but is in the spectroscopic 
sample of \citet{Beiler_2024}, who synthesize {\em JWST} colors. \citet{Beiler_2024} also provide a trigonometric parallax measurement for WISE-1047, whose discovery is presented in \citet{Meisner_2020b}. \citet{Beiler_2024} measure a parallax of $68.1 \pm 4.9$~mas, and calculate apparent magnitudes from their spectra of F150W $=$ 22.10 and F480M $=$ 16.33; the \href{https://jwst-docs.stsci.edu/jwst-calibration-status/nirspec-calibration-status/nirspec-fixed-slit-calibration-status#NIRSpecFixedSlitCalibrationStatus-Photometricrepeatability}{NIRSpec absolute calibration uncertainty} is estimated to be less than 3\% or 0.03 mag.

\loicadd{
Figure~\ref{fig:CMD} shows that WISE-1047 is blue or subluminous. \citet{Tu_2024} point out that the {\em JWST} spectra for this Y dwarf shows unusually strong CO$_2$ and CO absorption (see their Figure 4). 
Atmospheric models calculate that decreasing gravity results in increasing CO and CO$_2$ absorption and that these features are less sensitive to changes in metallicity\citep{
Karalidi_2021,Lacy_2023,Leggett_2024a}.  
The bandpass of the F480M filter directly samples the CO absorption, making it a useful indicator of the strength of this feature, an effect which could be masked by a broader filter.}

\loicadd{Figure~\ref{fig:CMD}
suggests log $g << 4.0$ and $T_{\rm eff} \approx 400$~K 
for WISE-1047. Evolutionary models \citep[e.g.][]{Marley_2021} then suggest a mass less than 3 Jupiter-masses and an age less than 0.2~Gyr for this Y dwarf.
Taking the proper motion for this source from \citet{Kirkpatrick_2021} and the parallax from \citet{Beiler_2024}, the BANYAN $\Sigma$ tool \citep{Gagne_2018} calculates a 52\% likelihood that WISE-1047 is a member of the Argus association, which has an age of $\sim 40~$Myr \citep{Zuckerman_2019}. If this brown dwarf is indeed this young, then it's mass is only 1 Jupiter-mass and log $g \approx 3.3$.  Further improvement to the distance measurement, and a radial velocity measurement, would be useful for this object.
}

\loicadd{\subsection{The Coldest Brown Dwarfs and Water Clouds}}

\loicadd{Water clouds are first expected to impact the photosphere when brown dwarfs cool to $T_{\rm eff} \sim 350$~K \citep[e.g.][]{Burrows_2003,Morley_2014, Lacy_2023}.} 
The \citet{Lacy_2023} disequilibrium chemistry sequences shown \loicadd{in Figure~\ref{fig:CMD}} are for clear atmospheres and for atmospheres with thin water clouds at pressures of 0.4 bar with particle size of 10\,$\mu$m (E10-type). The addition of clouds makes F150W brighter and F480M fainter \citep{Lacy_2023}, resulting in the
\loicadd{bluer F150W $-$ F480M colors}
seen in Figure~\ref{fig:CMD}. 

\loicadd{The coldest objects in our sample are 
WISE-0336B and WISE-0855, with $T_{\rm eff} \lesssim 300$~K.} The WISE-0336 binary components are separated by $0\farcs09$ or 1~au \citep{Calissendorff_2023}. 
The reanalysis of the photometry for the system presented here produces colors within 0.07 magnitudes of the \citet{Calissendorff_2023} 
values. Figure~\ref{fig:CMD} suggests that WISE-0336A
is a typical field Y dwarf with $T_{\rm eff}$ between 400~K and 450~K, 
\loicadd{and that WISE-0336B is slightly warmer than WISE-0855 but colder than the other Y dwarfs in the sample, with $T_{\rm eff}$ between 275~K and 300~K.} If the system has an age of $\sim 2$~Gyr \citep[e.g.][]{Best_2024}, then the component masses are approximately 12 and 5 Jupiter-masses \citep{Marley_2021}. 

Interestingly, WISE-0336B is significantly bluer than WISE-0855.  Given that WISE-0336A appears to have a gravity and metallicity typical of the field, the bluer color for 0336B is unlikely to be due to an unusually low gravity or high metallicity. 
Instead, the \citet{Lacy_2023} models suggest the difference is 
\loicadd{either that the atmosphere of WISE-0336B is cloudy and that of WISE-0855 is clear, or that WISE-0855 is cloudy but has a significantly higher gravity and/or lower metallicity than WISE-0336B. Recent analyses of {\em JWST} data for WISE-0855 support the former scenario, as \citet{Kuhnle_2024} find no
evidence of water clouds, and the metallicity and gravity appears typical of the field \citep{Luhman_2024,Kuhnle_2024}.}
\loicadd{The detection of water clouds may also be dependent on the surface gravity of the brown dwarf or on the viewing angle: observations of L dwarfs suggest} 
that the vertical extent of the water clouds is dependent on gravity \citep{Suarez_2023a, Suarez_2023b}, and the structured nature of surface storms and clouds results in colors being dependent on viewing angle \citep{Vos_2017}.


\bigskip

\bigskip
\section{Conclusions}

We share \loicadd{one of} the first homogeneous photometric datasets of Y-type brown dwarfs observed with {\it JWST}. 
The sample consists of 20 Y dwarfs observed with NIRCam in F150W and F480M, simultaneously, and 3 Y dwarfs observed with NIRISS in F480M. We present an F480M vs. F150W $-$ F480M color-magnitude diagram which confirms that Y dwarfs have a large, $\sim$1 mag, scatter at a given {\em near-infrared} $-~5~\mu$m color, as noted in previous studies. 

Of the different atmosphere models (Sonora Bobcat, Sonora Elf Owl, ATMO2020++ and Lacy \& Burrows), ATMO2020++ \loicadd{and  Lacy \& Burrows} best reproduce the colors of the 350~K -- 500~K Y dwarfs overall, while the Cloudy and Clear Lacy \& Burrows models reproduce the colors of the 300~K objects \loicadd{best}.  Both cloudy and clear atmospheres seem to be required to explain the observed colors of the coldest Y dwarfs.

We provide a photometry update for the Y+Y binary brown dwarf, WISE-0336A/B, and find that only the Cloudy Lacy \& Burrows model reproduces the colors of the faint secondary WISE-0336B.  WISE-0336B appears to be a cloudy version of WISE-0855, with $T_{\rm eff} \lesssim 300$~K.

We jointly present here and in Matuszewska et al. (in prep.) the first near-infrared detection of WISE-0830 and confirm its extremely red F150W $-$ F480M color (Figure 1, top left panel). 
The red color may be an indication of low metallicity and/or high gravity (i.e. and older-than-average age) for this high-velocity Y dwarf.

\loicadd{The F480M filter is particularly sensitive to the CO absorption band at $\lambda \approx 4.7~\mu$m. We find that one of the 400~K Y dwarfs in the \citet{Beiler_2024} sample, WISE-1047, is unusually blue in F150W $-$ F480M, and \citet{Tu_2024} notes that the spectrum of this brown dwarf has unusually strong CO (and CO$_2$) absorption. This spectral signature is indicative of low gravity according to atmospheric models, and the  BANYAN $\Sigma$ tool \citep{Gagne_2018} calculates a 52\% likelihood that WISE-1047 is a member of the $\sim$ 40~Myr-old Argus association; evolutionary models then imply that the object is extremely low-mass, only $\sim$ 1 Jupiter-mass.}

Finally, the astrometry at the {\it JWST} observation epoch is measured and we find agreement within \loicadd{2}$\sigma$ with the proper motion and parallax models of \cite{Kirkpatrick_2021} and Marocco et al. (in prep.), apart from one target\loicadd{, WISE-2354. The $2.8\sigma$ disagreement for WISE-2354 is nevertheless smaller than a NIRCam long-wave pixel.}   \loicadd{We updated the astrometry model of WISE-0830 by folding in our NIRCam measurement to reduce its parallax uncertainty by a factor of 2. This now puts WISE-0830 at a slightly smaller distance than previously measured of $10.1_{-0.6}^{+0.7}$\,pc.}

\begin{acknowledgments}

L.A. acknowledges support from the Canadian Space Agency funding program \# 22JWGO1-11. M.R.M., P.C., D.C.B.G., C.A.B., A.J.B., and J.A.G. gratefully acknowledge support through NASA grant Grant \# JWST-GO-02473.002-A. M.D.F. is supported by an NSF Astronomy and Astrophysics Postdoctoral Fellowship under award AST-2303911. T.R  would like to acknowledge support from NASA through the JWST NIRCam project through contract number NAS5-02105 (M. Rieke, University of Arizona, PI). CF acknowledges support from the Trottier Family Foundation and the Trottier Institute for Research on Exoplanets through her Trottier Postdoctoral Fellowship. Part of the work by C.A.B. was carried out at the Jet Propulsion Laboratory, California Institute of Technology, under a contract with the National Aeronautics and Space Administration (80NM0018D0004). This work was authored by A.Z.G. of Caltech/IPAC under Contract No. 80GSFC21R0032 with the National Aeronautics and Space Administration. This research made use of Photutils, an Astropy package for detection and photometry of astronomical sources \citep{photutils_1.8.0}. 

\end{acknowledgments}

\clearpage
\bibliography{bibliography}{}
\bibliographystyle{aasjournal}



\end{document}